\documentclass[a4paper,10pt]{article}
\usepackage{color}
\usepackage{cite}
\usepackage{amsmath,amsthm,amssymb,amsfonts}
\usepackage{eucal}
\usepackage[all]{xy}
\usepackage{color}
\usepackage{fancyhdr}
\usepackage{lastpage}
\usepackage{graphicx}
\usepackage{bm,bbm}
\usepackage{ascmac,paralist}
\usepackage{url}
\usepackage{array,arydshln}

\newtheorem{theorem}{Theorem}[section]

\newtheorem{corollary}[theorem]{Corollary}

\newtheorem{proposition}[theorem]{Proposition}

\newtheorem{example}[theorem]{Example}

\title{
Perfect state transfer, Equitable partition\\
and\\
Continuous-time quantum walk based search
}
\author{Yusuke Ide\thanks{Department of Mathematics, College of Humanities and Sciences, Nihon University, Setagaya, Tokyo, 156-8550, Japan, E-mail: ide.yusuke@nihon-u.ac.jp},
Akihiro Narimatsu\thanks{Center for Mathematical and Data Scienses, The University of Fukuchiyama, Fukuchiyama, Kyoto, 620-0886, Japan, E-mail: narimatsu-akihiro@fukuchiyama.ac.jp}
}
\date{}
\begin{document}
\maketitle
\begin{abstract}
In this paper, we consider a continuous-time quantum walk based search algorithm. We introduce equitable partition of the graph and perfect state transfer on it. By these two methods, we can calculate the success probability and the finding time of the search algorithm. In addition, we gave some examples of graphs that we can calculate the success probability and the finding time.

\end{abstract}
\medskip
\begin{flushleft}

\medskip
{\bf Keywords:} quantum walks, quantum search algorithms, perfect state transfer, equitable partition

\end{flushleft}
\section{Introduction}
The study of Quantum walks (QWs) has been paid much attention in the past two decades because of its applications\cite{kempe, kendon, konno, manouchehri, portugal, venegas}, especially in quantum information. The quantum search algorithm is one of the applications of QWs in quantum information \cite{ambaetal1, ambaetal2, chakraetal, childs, childsetal, childgold, loveetal, shenvietal}. 

In the previous study \cite{ide}, we proved that a graph partitioning method called {\it equitable partition} \cite{godroy} for the given graph reduce the size of the problem of the search algorithm. Under equitable partition, vertices of a graph are clustered and identified by distance from the marked vertex and edge connection with other vertices. We should remark that the initial state of the search algorithm is uniform state. Although the initial state of the search algorithm is uniform, the difference of the numbers of vertices in each partition causes non-uniform effects for the uniform initial state.

In this paper, we introduce {\it perfect state transfer} (PST) which is a state transfer phenomenon related to quantum walks. 
PST has been actively studied in mathematics \cite{godsil1, kubotasegawa, sabri} and physics \cite{akay}. In mathematics, there are many studies from the viewpoint of spectral graph theory and several properties are revealed \cite{coutinho}. For example, when PST from a vertex $u$ to a vertex $v$ at time $\tau$ occurs, PST from the vertex $v$ to the vertex $u$ at time $\tau$ also occurs. Considering this under equitable partition, we obtain our main result about quantum search algorithm. In addition, we propose some meaningful example of the theorem as follows. Let $G=(V(G),E(G))$ be a simple graph with its vertex set $V(G)=\{0,1,\dots,N-1\}$ and edge set $E(G)\subset V(G)\times V(G)$. The matrix $\overline{H}_G$ is defined in section $2.3$, Eq.\eqref{hgbarag}.

\begin{example}
When $\overline{H}_G$ is given by
\begin{align*}
\overline{H}_G=| 0\rangle\langle 0|+\frac{1}{N-2}
\begin{bmatrix}
0&\sqrt{N-1}\\
\sqrt{N-1}&N-2
\end{bmatrix}
=\begin{bmatrix}
1&\frac{\sqrt{N-1}}{N-2}\\
\frac{\sqrt{N-1}}{N-2}&1
\end{bmatrix},
\end{align*}
we have the success probability of the search algorithm $\mathbb{P}_{N,T}(0)$ at time \\$\displaystyle{T=\frac{(N-2)\pi}{2\sqrt{N-1}}={\mathcal O}(\sqrt{N})}$ is
\begin{align*}
\mathbb{P}_{N,T}(0)=1-\frac{1}{N}.
\end{align*}
This situation corresponds to the case where $G$ is a complete graph, $G_{\overline{0}}=(\{0\},\emptyset)$ is a graph consists of the marked one vertex $0$, $G_{\overline{1}}$ is a complete graph with $N-1$ vertices and the partition $(G_{\overline{0}},G_{\overline{1}})$ is equitable partition of $G$. 
\end{example}

\begin{example}
When $\overline{H}_G$ is given by
\begin{align*}
\overline{H}_G&=| 0\rangle\langle 0|+\frac{1}{2}
\begin{bmatrix}
0&(2k+1)\sqrt{2}&0\\
(2k+1)\sqrt{2}&2&(2k+1)\sqrt{2}\\
0&(2k+1)\sqrt{2}&2
\end{bmatrix}\\
&=\begin{bmatrix}
1&\frac{(2k+1)\sqrt{2}}{2}&0\\
\frac{(2k+1)\sqrt{2}}{2}&1&\frac{(2k+1)\sqrt{2}}{2}\\
0&\frac{(2k+1)\sqrt{2}}{2}&1
\end{bmatrix},
\end{align*}
where $k$ is a positive integer, we have the success probability of the search algorithm $\mathbb{P}_{N,T}(0)$ at time $T=\pi$,
\begin{align*}
\mathbb{P}_{N,T}(0)\approx 1-\frac{1}{\sqrt{N}}.
\end{align*}
This situation corresponds to the case where $G_{\overline{0}}=(\{0\},\emptyset)$ is a graph consists of the marked one vertex $0$, $G_{\overline{1}}$ is a cycle graph with $2(2k+1)^2$ vertices, $G_{\overline{2}}$ is a cycle graph with $\{2(2k+1)^2\}^2$ vertices. 
\end{example}

The rest of this paper is organized as follows: Section 2 is devoted to the definition of our model and the introduction of a related result. Section 3 gives our result and introduction of the examples. Section 4 summerizes our paper. 

\section{Preliminaries}
\subsection{Continuous-time quantum walk}
Let $G=(V(G),E(G))$ be a simple undirected graph without self-loops and multiple edges, with its vertex set $V(G)=\{0,1,\dots,N-1\}$ and edge set $E(G)\subset V(G)\times V(G)$. As an example, for the complete graph on $N$ vertices $K_N$, which is the simple graph with $\begin{pmatrix}N\\2\end{pmatrix}$ edges and the edge set is defined by $E(K_N)=\{(j,k)\ :\ 0\leq j<k\leq N-1\}$. Continuous-time quantum walk (CTQW) is a unitary time evolution process on the graph with the Hilbert space,
\begin{align*}
{\mathcal H}={\rm Span}\{|j\rangle:j\in V(G)\}.
\end{align*}
On this Hilbert space, we define the unitary time evolution operator of CTQW. We choose an $N$-dimensional Hermitian matrix $M_G$ with $(M_G)_{j,k}\neq 0$ when $(j,k)\in E(G)$ and $(M_G)_{j,k}= 0$ when $(j,k)\notin E(G)$. The component $(M_G)_{j,k}$ is regarded as ``the weight of the edge $(j,k)\in E(G)$''. One of the typical $M_G$ is the adjacency matrix
\begin{align*}
(A_G)_{j,k}=\begin{cases}
1\quad (j,k)\in E(G),\\
0\quad {\rm otherwise.}
\end{cases}
\end{align*}
of the graph $G$.

For the time $t\geq0$, we define the time evolution operator
\begin{align}
U_{M_G}(t)=\exp(itM_G)=\sum_{k=0}^{\infty}\frac{(it)^k}{k !}M_G^k, \label{timeevo}
\end{align}
of CTQW on $G$ corresponding to $M_G$. We should note that CTQW is a quantum dynamics determined by Schr\"{o}dinger equation with its Hamiltonian $M_G$. We set a vector $|\Psi_{N,0}\rangle\in {\mathcal H}(\| |\Psi_{N,0}\rangle\|=1)$ as initial state. Then the time evolution is written by
\begin{align*}
|\Psi_{N,t}\rangle=U_{M_G}(t)|\Psi_{N,0}\rangle,
\end{align*}
for $t\geq 0$. $|\Psi_{N,t}\rangle$ describes the probability amplitude at time $t$. We put the finding probability of the quantum walker on $s\in V(G)$ at time $t$ as
\begin{align}
\mathbb{P}_{N,t}(s)=|\langle s|\Psi_{N,t}\rangle|^2.\label{prob}
\end{align}
\subsection{The quantum walk based search algorithm}
In this subsection, we consider CTQW based search on simple graph $G=(V(G),E(G))$ with $V(G)=\{0,1,\dots,N-1\}$. The Hermitian matrix $H_G$ for CTQW search on $G$ is defined by
\begin{align*}
H_G=|w\rangle\langle w|+\gamma A_G,
\end{align*}
where $w\in V(G)$ is the marked vertex. Now we consider the CTQW on $G$ with its time evolution operator $U_{H_G}$. The main task for CTQW search is finding a suitable $\gamma\in\mathbb{R}$ and a time $T\leq{\mathcal O}(\sqrt{N})$ such that we can attain
\begin{align*}
\mathbb{P}_{N,T}(w)\approx1\quad(N\to \infty),
\end{align*}
with the uniform initial state
\begin{align*}
|\Psi_0\rangle=\frac{1}{\sqrt{N}}\sum_{j=0}^{N-1}|j\rangle.
\end{align*}
\subsection{Equitable partition}
This subsection is a review of \cite{ide}, section $4$ and deals with a partition of the graph, called equitable partition for CTQW search. Without loss of generality, we assume that the marked vertex is $0\in V(G)$. We consider the partition $(G_{\overline{0}},G_{\overline{1}},\dots,G_{\overline{J-1}})$ of $G=(V(G),E(G))$ which satisfies the following $3$ conditions:
\begin{enumerate}
\item $V(G_{\overline{0}})=\{0\}$.
\item $V(G)=\bigcup_{\overline{j}=0}^{J-1}V(G_{\overline{j}}),$\quad $V(G_{\overline{j}})\cap V(G_{\overline{k}})=\emptyset(\overline{j}\neq\overline{k})$.
\item For each $\overline{0}\leq \overline{j},\overline{k}\leq\overline{J-1}$, there exists a non-negative integer $d_{\overline{j},\overline{k}}$ such that $d_{\overline{j},\overline{k}}=|\{w\in V(G_{\overline{k}}):(v,w)\in E(G)\}|$ for each $v\in V(G_{\overline{J}})$.
\end{enumerate}
We use the notation $G_{\overline{j}}\sim G_{\overline{k}}$ if $d_{\overline{j},\overline{k}}\neq 0$.

The partition $(G_{\overline{0}},G_{\overline{1}},\dots,G_{\overline{J-1}})$ of $G=(V(G),E(G))$ consists of $\overline{J}$ subgraphs. For each subgraph $G_{\overline{j}}$ is $d_{\overline{j},\overline{j}}$-regular graph. Especially, $G_{\overline{0}}$ consists of only the marked vertex $0\in V(G)(d_{\overline{0},\overline{0}}=0)$, which is called the null graph. In addition, each of the vertices $v\in V(G_{\overline{j}})$ connected to the same number $d_{\overline{j},\overline{k}}$ of the vertices in $V(G_{\overline{k}})$.

From now on, we consider the CTQW search on simple and connected graph $G$ with equitable partition $(G_{\overline{0}},G_{\overline{1}},\dots,G_{\overline{J-1}})$ with $n_{\overline{j}}=|V(G_{\overline{j}})|$ of $G$. We define the uniform states related to the partition as follows.
\begin{align*}
|\overline{j}\rangle=\frac{1}{\sqrt{n_{\overline{j}}}}\sum_{j\in V(G_{\overline{j}})}|j\rangle,
\end{align*}
for $0\leq \overline{j}\leq J-1$. By direct calculation, we have
\begin{align*}
A_G|\overline{j}\rangle&=d_{\overline{j},\overline{j}}|\overline{j}\rangle+\sum_{G_{\overline{j}}\sim G_{\overline{k}}}d_{\overline{k},\overline{j}}\sqrt{\frac{n_{\overline{k}}}{n_{\overline{j}}}}|\overline{k}\rangle=d_{\overline{j},\overline{j}}|\overline{j}\rangle+\sum_{G_{\overline{j}}\sim G_{\overline{k}}}\sqrt{d_{\overline{k},\overline{j}}}\sqrt{\frac{n_{\overline{k}}d_{\overline{j},\overline{k}}}{n_{\overline{j}}}}|\overline{k}\rangle\\
&=d_{\overline{j},\overline{j}}|\overline{j}\rangle+\sum_{G_{\overline{j}}\sim G_{\overline{k}}}\sqrt{d_{\overline{j},\overline{k}}d_{\overline{k},\overline{j}}}|\overline{k}\rangle\\
&=\sum_{\overline{0}\leq \overline{J-1}}\sqrt{d_{\overline{j},\overline{k}}d_{\overline{k},\overline{j}}}|\overline{k}\rangle,
\end{align*}
where $n_{\overline{j}}d_{\overline{j},\overline{k}}=n_{\overline{k}}d_{\overline{k},\overline{j}}$ coming from simplicity of the graph. Here we define a $J\times J$ matrix $\overline{A}_G$ as
\begin{align}
(\overline{A}_G)_{\overline{j},\overline{k}}=\sqrt{d_{\overline{j},\overline{k}}d_{\overline{k},\overline{j}}}.\label{agbar}
\end{align}
When we define a $J\times J$ matrix $\overline{H}_G$ as
\begin{align}
\overline{H}_G={\rm diag}(1,0,\dots,0)+\gamma\overline{A}_G,\label{hgbarag}
\end{align}
the action of the Hermitian matrix $H_G$ is closed on the subspace Span$\{|\overline{j}\rangle:\overline{0}\leq\overline{j}\leq\overline{J-1}\}$ as follows.
\begin{align*}
H_G|\overline{j}\rangle=\overline{H}_G|\overline{j}\rangle,
\end{align*}
for $\overline{0}\leq\overline{j}\leq\overline{J-1}$. We also have $\displaystyle{|\Psi_0\rangle=\frac{1}{\sqrt{N}}\sum_{\overline{j}=\overline{0}}^{\overline{J-1}}\sqrt{n_{\overline{j}}}|\overline{j}\rangle}\in$Span$\{|\overline{j}\rangle:\overline{0}\leq\overline{j}\leq\overline{J-1}\}$. Then we obtain
\begin{align*}
H_G|\Psi_0\rangle=\overline{H}_G\Biggl(\frac{1}{\sqrt{N}}\sum_{\overline{j}=\overline{0}}^{\overline{J-1}}\sqrt{n_{\overline{j}}}|\overline{j}\rangle \Biggr),
\end{align*}
thus we have
\begin{align}
{\rm exp}(itH_G)|\Psi_0\rangle={\rm exp}(it\overline{H}_G)\Biggl(\frac{1}{\sqrt{N}}\sum_{\overline{j}=\overline{0}}^{\overline{J-1}}\sqrt{n_{\overline{j}}}|\overline{j}\rangle \Biggr).\label{exph-hbar}
\end{align}

\subsection{Perfect state transfer}
In this subsection, we consider PST related to $M_G$. PST related to $M_G$ from a vertex $j$ to a vertex $k$ at time $\tau$ with phase $\lambda$ occurs if and only if
\begin{align}
\bigl(U_{M_G}(\tau)\bigr)_{k,j}=\lambda, \label{pstdef}
\end{align}
where $j,k\in \{0,1,\dots,N-1\}$, $|\lambda|=1$. Combining Eq.\eqref{timeevo} with Eq.\eqref{pstdef} gives 
\begin{align*}
\langle j|\Psi_{N,0}\rangle=\lambda\langle k|\Psi_{N,\tau}\rangle.
\end{align*}
Noting $|\lambda|=1$, this equation implies that the state on a vertex $j$ moves to a vertex $k$ after time $\tau$. 

Since $M_G$ be a Hermitian matrix, eigenvalues of $M_G$ and the each component of the corresponding eigenvectors can be real numbers. Let $\theta_0\geq\theta_1\geq\dots\geq\theta_{N-1}$ be eigenvalues of $M_G$ and $|v_0\rangle,|v_1\rangle,\dots,|v_{N-1}\rangle$ be corresponding ${\mathbb R}^N$ valued eigenvectors. 
Previous study \cite{coutinho, godsil1} revealed the following claim. 
\begin{proposition}
A necessary and sufficient condition that the time evolution operator must satisfy for perfect State Transfer between a vertex $j$ and a vertex $k$ at time $\tau$ with phase $\lambda$ to occur is following three properties.
\begin{enumerate}
\item For all $\ell\in\{0,\dots,N-1\}$, eigenvalues $\theta_\ell$ and corresponding eigenvectors $|v_\ell\rangle$ satisfy $\langle v_\ell|j\rangle=\pm\langle v_\ell|k\rangle$.
\item If $\langle v_\ell|j\rangle=\langle v_\ell|k\rangle$, there exist $n\in\mathbb{Z}$ such that $\theta_0-\theta_\ell=2n\pi$.
\item If $\langle v_\ell|j\rangle=-\langle v_\ell|k\rangle$, there exist $n\in\mathbb{Z}$ such that $\theta_0-\theta_\ell=(2n+1)\pi$.
\end{enumerate}
Under these conditions, $\lambda=e^{i\tau \theta_0}.$
\end{proposition}
Using this proposition, the three properties can be rewritten as given in appendix. 

\section{Result}
This section is devoted to our theorem and  some examples refered in Section $1$. 

\subsection{Theorem}
\begin{theorem}
If perfect state transfer between vertex $\overline{0}$ and a vertex $\overline{j}$ at time $\tau$ with phase $\lambda$ occurs for CTQW using $\overline{H}_{G}$, success probability of the search algorithm is given by
\begin{align*}
\mathbb{P}_{N,\tau}(0)=\frac{n_{\overline{j}}}{N}.
\end{align*}
\end{theorem}
\noindent
{\it proof.} We should note that the initial state $|\Psi_0\rangle$ can be written as
\begin{align*}
|\Psi_0\rangle=\frac{1}{\sqrt{N}}\sum_{\overline{j}=\overline{0}}^{\overline{J-1}}\sqrt{n_{\overline{j}}}|\overline{j}\rangle.
\end{align*}
Combining Eq.\eqref{prob} with Eq.\eqref{exph-hbar} gives
\begin{align*}
\mathbb{P}_{N,\tau}(0)&=\bigl|\langle 0|\exp(i\tau H_G)|\Psi_0\rangle\bigr|^2\\
&=\Biggl|\langle \overline{0}|\exp(i\tau \overline{H}_G)\Biggl(\sum_{\overline{k}=\overline{0}}^{\overline{J-1}}\frac{\sqrt{n_{\overline{k}}}}{\sqrt{N}}|\overline{k}\rangle \Biggr)\Biggr|^2.
\end{align*}
Since perfect state transfer between $\overline{0}$ and $\overline{j}$ occur, we have
\begin{align*}
\mathbb{P}_{N,\tau}(0)&=\Biggl|\langle \overline{0}|\lambda\frac{\sqrt{n_{\overline{j}}}}{\sqrt{N}}|\overline{0}\rangle \Biggr|^2
=\Biggl|\lambda\sqrt{\frac{n_{\overline{j}}}{N}}\Biggr|^2
=\frac{n_{\overline{j}}}{N}.
\end{align*}
\hfill$\square$
\subsection{Examples}
\setcounter{section}{1}
\setcounter{theorem}{0}
In this subsection, we consider examples $1.1.$ and $1.2.$, which are the examples of our theorem. 
\begin{example}
When $\overline{H}_G$ is given by
\begin{align*}
\overline{H}_G=| 0\rangle\langle 0|+\frac{1}{N-2}
\begin{bmatrix}
0&\sqrt{N-1}\\
\sqrt{N-1}&N-2
\end{bmatrix}
=\begin{bmatrix}
1&\frac{\sqrt{N-1}}{N-2}\\
\frac{\sqrt{N-1}}{N-2}&1
\end{bmatrix},
\end{align*}
we have the success probability of the search algorithm $\mathbb{P}_{N,T}(0)$ at time \\
$\displaystyle{T=\frac{(N-2)\pi}{2\sqrt{N-1}}={\mathcal O}(\sqrt{N})}$ is
\begin{align*}
\mathbb{P}_{N,T}(0)=1-\frac{1}{N}.
\end{align*}
This situation corresponds to the case where $G$ is a complete graph, $G_{\overline{0}}=(\{0\},\emptyset)$ is a graph consists of the marked vertex $0$, $G_{\overline{1}}$ is a complete graph with $N-1$ vertices and the partition $(G_{\overline{0}},G_{\overline{1}})$ is equitable partition of $G$. 
\end{example}
\noindent
{\it proof.} By Eqs.\eqref{agbar} and \eqref{hgbarag}, we have $\gamma=1/(N-2)$ and 
\begin{align*}
\overline{A}_G&=
\begin{bmatrix}
0&\sqrt{N-1}\\
\sqrt{N-1}&N-2
\end{bmatrix}
=\begin{bmatrix}
d_{\overline{0},\overline{0}}&\sqrt{d_{\overline{0},\overline{1}}d_{\overline{1},\overline{0}}}\\
\sqrt{d_{\overline{1},\overline{0}}d_{\overline{0},\overline{1}}}&d_{\overline{1},\overline{1}}
\end{bmatrix}.
\end{align*}
Since $d_{\overline{0},\overline{0}}=0$, we have $|V(G_{\overline{0}})|=1$. Thus we get $d_{\overline{0},\overline{1}}=N-1$, $|V(G_{\overline{1}})|=N-1$ and $G_{\overline{1}}$ is a complete graph with $N-1$ vertices. Then we obtain $G$ is a complete graph with $N$ vertices.

Next we consider the success probability. Let $\theta_{\pm}$ be eigenvalues of $\overline{H}_G$ and $|v_{\pm}\rangle$ be corresponding eigenvectors. By spectral decomposition, we have
\begin{align*}
\overline{H}_{G}&=\theta_{+}|v_{+}\rangle\langle v_{+}|+\theta_{-}|v_{-}\rangle\langle v_{-}| \\
&=\theta_{+}\frac{1}{\sqrt{2}}\begin{bmatrix}1\\1\end{bmatrix}\frac{1}{\sqrt{2}}\begin{bmatrix}1&1\end{bmatrix}
+\theta_{-}\frac{1}{\sqrt{2}}\begin{bmatrix}1\\-1\end{bmatrix}\frac{1}{\sqrt{2}}\begin{bmatrix}1&-1\end{bmatrix}\\
&=\frac{\theta_{+}}{2}\begin{bmatrix}1&1\\1&1\end{bmatrix}+\frac{\theta_{-}}{2}\begin{bmatrix}1&-1\\-1&1\end{bmatrix} ,
\end{align*}
where
\begin{align*}
\theta_{\pm}=1\pm\frac{\sqrt{N-1}}{N-2}.
\end{align*}
Then we obtain
\begin{align*}
e^{it\overline{H}_G}&=
e^{it\theta_{+}}\frac{1}{2}
\begin{bmatrix}
1&1\\
1&1
\end{bmatrix}
+e^{it\theta_{-}}\frac{1}{2}
\begin{bmatrix}
1&-1\\
-1&1
\end{bmatrix}\\
&=e^{it}
\begin{bmatrix}
\cos\Bigl(\frac{\sqrt{N-1}}{N-2}t\Bigr)&i\sin\Bigl(\frac{\sqrt{N-1}}{N-2}t\Bigr)\\
i\sin\Bigl(\frac{\sqrt{N-1}}{N-2}t\Bigr)&\cos\Bigl(\frac{\sqrt{N-1}}{N-2}t\Bigr)
\end{bmatrix}.
\end{align*}
Therefore when time $\displaystyle{T=\frac{(N-2)\pi}{2\sqrt{N-1}}={\mathcal O}(\sqrt{N})}$, 
\begin{align*}
e^{iT\overline{H}_G}=e^{iT}\begin{bmatrix}
\cos\Bigl(\frac{\sqrt{N-1}}{N-2}T\Bigr)&i\sin\Bigl(\frac{\sqrt{N-1}}{N-2}T\Bigr)\\
i\sin\Bigl(\frac{\sqrt{N-1}}{N-2}T\Bigr)&\cos\Bigl(\frac{\sqrt{N-1}}{N-2}T\Bigr)
\end{bmatrix}
=e^{iT}\begin{bmatrix}
0&i\\
i&0
\end{bmatrix}.
\end{align*}
Thus PST occurs from $\overline{1}$ to $\overline{0}$ at time $\displaystyle{T=\frac{(N-2)\pi}{2\sqrt{N-1}}}$. Then the finding probability is given by
\begin{align*}
\mathbb{P}_{N,T}(0)=\frac{n_{\overline{1}}}{N}=\frac{N-1}{N}=1-\frac{1}{N}.
\end{align*}
\hfill$\square$
\begin{example}
When $\overline{H}_G$ is given by
\begin{align*}
\overline{H}_G&=| 0\rangle\langle 0|+\frac{1}{2}
\begin{bmatrix}
0&(2k+1)\sqrt{2}&0\\
(2k+1)\sqrt{2}&2&(2k+1)\sqrt{2}\\
0&(2k+1)\sqrt{2}&2
\end{bmatrix}\\
&=\begin{bmatrix}
1&\frac{(2k+1)\sqrt{2}}{2}&0\\
\frac{(2k+1)\sqrt{2}}{2}&1&\frac{(2k+1)\sqrt{2}}{2}\\
0&\frac{(2k+1)\sqrt{2}}{2}&1
\end{bmatrix},
\end{align*}
where $k$ is a positive integer, we have the success probability of the search algorithm $\mathbb{P}_{N,T}(0)$ at time $T=\pi$,
\begin{align*}
\mathbb{P}_{N,T}(0)\approx 1-\frac{1}{\sqrt{N}}.
\end{align*}
This situation corresponds to the case where $G_{\overline{0}}=(\{0\},\emptyset)$ is a graph consists of the marked one vertex $0$, $G_{\overline{1}}$ is a cycle graph with $2(2k+1)^2$ vertices, $G_{\overline{2}}$ is a cycle graph with $\{2(2k+1)^2\}^2$ vertices. 
\end{example}
\noindent
{\it proof.} By Eqs.\eqref{agbar} and \eqref{hgbarag}, we have $\gamma=1/2$ and
\begin{align*}
\overline{A}_G&=\begin{bmatrix}
0&(2k+1)\sqrt{2}&0\\
(2k+1)\sqrt{2}&2&(2k+1)\sqrt{2}\\
0&(2k+1)\sqrt{2}&2
\end{bmatrix}\\
&=\begin{bmatrix}
d_{\overline{0},\overline{0}}&\sqrt{d_{\overline{0},\overline{1}}d_{\overline{1},\overline{0}}}&\sqrt{d_{\overline{0},\overline{2}}d_{\overline{2},\overline{0}}}\\
\sqrt{d_{\overline{1},\overline{0}}d_{\overline{0},\overline{1}}}&d_{\overline{1},\overline{1}}&\sqrt{d_{\overline{1},\overline{2}}d_{\overline{2},\overline{1}}}\\
\sqrt{d_{\overline{2},\overline{0}}d_{\overline{0},\overline{2}}}&\sqrt{d_{\overline{2},\overline{1}}d_{\overline{1},\overline{2}}}&d_{\overline{2},\overline{2}}
\end{bmatrix}.
\end{align*}
There are some choices for each $d_{\overline{j},\overline{k}}$. We put
\begin{align*}
\begin{bmatrix}
d_{\overline{0},\overline{0}}&d_{\overline{0},\overline{1}}&d_{\overline{0},\overline{2}}\\
d_{\overline{1},\overline{0}}&d_{\overline{1},\overline{1}}&d_{\overline{1},\overline{2}}\\
d_{\overline{2},\overline{0}}&d_{\overline{2},\overline{1}}&d_{\overline{2},\overline{2}}
\end{bmatrix}=
\begin{bmatrix}
0&2(2k+1)^2&0\\1&2&2(2k+1)^2\\0&1&2
\end{bmatrix}.
\end{align*}
$d_{\overline{0},\overline{0}}=0$ implies that $|V(G_{\overline{0}})|=1$, or in other words, $G_{\overline{0}}$ consists of only one marked vertex $0$. By $d_{\overline{1},\overline{1}}=2$ and $d_{\overline{2},\overline{2}}=2$, $G_{\overline{1}}$ and $G_{\overline{2}}$ are $2$-regular graph. We chose the cycle graph for $G_{\overline{1}}$ and $G_{\overline{2}}$ because the cycle graph is a well known example of $2$-regular graphs. Since $d_{\overline{0},\overline{1}}=2(2k+1)^2$ and $V(G_{\overline{0}})=\{0\}$, $|V(G_{\overline{1}})|=2(2k+1)^2$. Therefore, $G_{\overline{1}}$ is a cycle graph with $2(2k+1)^2$ vertices. $d_{\overline{1},\overline{2}}=2(2k+1)^2$ and $d_{\overline{2},\overline{1}}=1$, each vertex in $G_{\overline{1}}$ is connected to $2(2k+1)^2$ different vertices in $G_{\overline{2}}$ and each vertex in  $G_{\overline{2}}$ is connected to $1$ vertex in $G_{\overline{1}}$. Thus we obtain $|V(G_{\overline{2}})|=\{2(2k+1)^2\}^2$ and $G_{\overline{2}}$ is a cycle graph with $\{2(2k+1)^2\}^2$ vertices. 

Next we consider the success probability. Let $\theta_{0},\theta_{1}, \theta_{2}$ be eigenvalues of $\overline{H}_G$ and $|v_{0}\rangle,|v_{1}\rangle,|v_{2}\rangle$ be corresponding eigenvectors. By spectral decomposition, we obtain 
\begin{align*}
\overline{H}_G&=\theta_{0}|v_{0}\rangle\langle v_{0}|+\theta_{1}|v_{1}\rangle\langle v_{1}|+\theta_{2}|v_{2}\rangle\langle v_{2}|\\
&=\theta_{0}\frac{1}{2}\begin{bmatrix}1\\ \sqrt{2}\\1\end{bmatrix}\frac{1}{2}\begin{bmatrix}1& \sqrt{2}&1\end{bmatrix} +\theta_{1}\frac{1}{\sqrt{2}}\begin{bmatrix}1\\ 0\\-1\end{bmatrix}\frac{1}{\sqrt{2}}\begin{bmatrix}1& 0&-1\end{bmatrix}\\&+\theta_{2}\frac{1}{2}\begin{bmatrix}1\\ -\sqrt{2}\\1\end{bmatrix}\frac{1}{2}\begin{bmatrix}1& -\sqrt{2}&1\end{bmatrix}\\
&=\frac{\theta_{0}}{4}\begin{bmatrix}
1&\sqrt{2}&1\\ \sqrt{2}&2&\sqrt{2}\\ 1&\sqrt{2}&1
\end{bmatrix}
+\frac{\theta_{1}}{2}\begin{bmatrix}
1&0&-1\\0&0&0\\-1&0&1
\end{bmatrix}
+\frac{\theta_{2}}{4}\begin{bmatrix}
1&-\sqrt{2}&1\\ -\sqrt{2}&2&-\sqrt{2}\\ 1&-\sqrt{2}&1
\end{bmatrix},
\end{align*}
where $\theta_{0}=2(k+1),\theta_{1}=1,\theta_{2}=-2k,$ with $k$ is a positive integer. Then we obtain
\begin{align*}
e^{iT\overline{H}_G}&=e^{iT\theta_{0}}\frac{1}{4}\begin{bmatrix}
1&\sqrt{2}&1\\ \sqrt{2}&2&\sqrt{2}\\ 1&\sqrt{2}&1
\end{bmatrix}
+e^{iT\theta_{1}}\frac{1}{2}\begin{bmatrix}
1&0&-1\\0&0&0\\-1&0&1
\end{bmatrix}\\
&+e^{iT\theta_{2}}\frac{1}{4}\begin{bmatrix}
1&-\sqrt{2}&1\\ -\sqrt{2}&2&-\sqrt{2}\\ 1&-\sqrt{2}&1
\end{bmatrix}.
\end{align*}
When $T=\pi$, we obtain
\begin{align*}
e^{i\pi \overline{H}_G}&=e^{i\pi 2(k+1)}\frac{1}{4}\begin{bmatrix}
1&\sqrt{2}&1\\ \sqrt{2}&2&\sqrt{2}\\ 1&\sqrt{2}&1
\end{bmatrix}
+e^{i\pi\cdot 1}\frac{1}{2}\begin{bmatrix}
1&0&-1\\0&0&0\\-1&0&1
\end{bmatrix}\\
&+e^{i\pi(-2k)}\frac{1}{4}\begin{bmatrix}
1&-\sqrt{2}&1\\ -\sqrt{2}&2&-\sqrt{2}\\ 1&-\sqrt{2}&1
\end{bmatrix}=\begin{bmatrix}
0&0&1\\0&1&0\\1&0&0
\end{bmatrix}.
\end{align*}
Thus PST occurs from $\overline{2}$ to $\overline{0}$ at time $T=\pi$. Then the finding probability is given by
\begin{align*}
{\mathbb P}_{N,\pi}(0)=\frac{n_{\overline{2}}}{N}=\frac{\{2(2k+1)^2\}^2}{\{2(2k+1)^2\}^2+2(2k+1)^2+1}\approx 1-\frac{1}{\sqrt{N}}.
\end{align*}
\hfill$\square$

When $k=1$, Example $1.2.$ includes the case in \cite{godsil1}.
\setcounter{section}{3}
\section{Summary}
In this paper, we analyzed the continuous-time quantum walk based search algorithm. We obtained the success probability of the search algorithm by using PST on equitable partition of the graph. In addition, we proposed some graphs as examples of our theorem. One of the interesting future problem is to reduce the conditions of equitable partition. 
\newpage
\appendix
\section{Appendix}
Let each component of the eigenvector of Hermitian matrix $M_G$ be
\begin{align*}
|v_\ell\rangle=\begin{bmatrix}
|v_\ell(0)|e^{\Tilde{\theta}_{\ell}(0)}\\
|v_\ell(1)|e^{\Tilde{\theta}_{\ell}(1)}\\
\vdots\\
|v_\ell(N-1)|e^{\Tilde{\theta}_{\ell}(N-1)}
\end{bmatrix}.
\end{align*}
\begin{corollary}
A necessary and sufficient condition that the time evolution operator must satisfy for PST from a vertex $j$ to a vertex $k$ at time $\tau$ to occur is following.
\begin{align*}
(\theta_0-\theta_\ell)\tau\equiv \Tilde{\theta}_{\ell}(k)-\Tilde{\theta}_{\ell}(j)-\{\Tilde{\theta}_{0}(k)-\Tilde{\theta}_{0}(j)\}.\quad (\bmod \ 2\pi)
\end{align*}
\end{corollary}
\noindent
{\it proof.} By Eq.\eqref{pstdef}, we have
\begin{align*}
U_{M_G}(\tau)|j\rangle=\lambda|k\rangle.
\end{align*}
By direct calculation, we get
\begin{align*}
\Biggl(\sum_{\ell=0}^{N-1}e^{i\tau\theta_{\ell}}|v_\ell\rangle\langle v_\ell|\Biggr)|j\rangle=\lambda\Biggl(\sum_{\ell=0}^{N-1}|v_\ell\rangle\langle v_\ell|\Biggr)|k\rangle,
\end{align*}
comparing each component of the left hand side with that of the right hand side, we obtain
\begin{align*}
e^{i\tau\theta_{\ell}}\langle v_{\ell}|j\rangle=\lambda\langle v_{\ell}|k\rangle\quad({\rm for\ all\ }\ell=0,\dots,N-1),
\end{align*}
computing this, we have
\begin{align*}
e^{i\tau\theta_{\ell}}|v_{\ell}(j)|e^{-i\tilde{\theta}_{\ell}(j)}=\lambda|v_{\ell}(k)|e^{-i\tilde{\theta}_{\ell}(k)}\quad({\rm for\ all\ }\ell=0,\dots,N-1).
\end{align*}
Noting that $|\lambda|=1$, we get
\begin{align}
\begin{cases}
e^{i\tau\theta_{\ell}}e^{-i\tilde{\theta}_{\ell}(j)}=\lambda e^{-i\tilde{\theta}_{\ell}(k)}\\
|v_{\ell}(j)|=|v_{\ell}(k)|.
\end{cases}\label{coroa11}
\end{align}
When $\ell=0$, Eq.\eqref{coroa11} gives
\begin{align}
\lambda=e^{i\tau\theta_{0}}e^{i(\tilde{\theta}_0(k)-\tilde{\theta}_0(j))}\label{coroa12}
\end{align}
Substituting Eq.\eqref{coroa12} into Eq.\eqref{coroa11}, we obtain
\begin{align*}
e^{i(\tilde{\theta}_{\ell}(k)-\tilde{\theta}_{\ell}(j))}=e^{i\tau(\theta_{0}-\theta_{\ell})}e^{i(\tilde{\theta}_0(k)-\tilde{\theta}_0(j))}
\end{align*}
calculating this, we have
\begin{align*}
e^{i\{(\tilde{\theta}_{\ell}(k)-\tilde{\theta}_{\ell}(j))-(\tilde{\theta}_0(k)-\tilde{\theta}_0(j))\}}=e^{i\tau(\theta_{0}-\theta_{\ell})}.
\end{align*}
Thus we get
\begin{align*}
(\theta_0-\theta_\ell)\tau\equiv \Tilde{\theta}_{\ell}(k)-\Tilde{\theta}_{\ell}(j)-\{\Tilde{\theta}_{0}(k)-\Tilde{\theta}_{0}(j)\}.\quad (\bmod \ 2\pi)
\end{align*}
\hfill$\square$

\end{document}